# Constructed Realities? Technical and Contextual Anomalies in a High-Profile Image

Matthias Wjst[1]

[1] contact address: Institut für KI und Informatik in der Medizin, Lehrstuhl für Medizinische Informatik, Klinikum rechts der Isar, Grillparzerstr. 18, D-81675 München, Germany, wjst@tum.de
http://orcid.org/0000-0002-4974-5631

1/5/2026

Abstract

This study offers a forensic assessment of a widely circulated photograph featuring Andrew Mountbatten-Windsor, Virginia Giuffre, and Ghislaine Maxwell, an image that has played a pivotal role in public discourse and legal narratives. Numerous inconsistencies emerge across multiple published versions, including irregularities in lighting, posture, and physical interaction, which are more consistent with digital compositing than with an unmanipulated original. The analysis includes a 3D reconstruction of the scene geometry and a search of reference images matched to the identified camera model. Because no original print is available, and no verifiable chain of custody exists for the original, definitive conclusions remain unattainable. Even so, the technical and contextual anomalies indicate that the photograph may have been deliberately fabricated, particularly since two probable source image unrelated to the case were identified. In the absence of further evidence, it remains an unresolved yet symbolically charged artifact within a complex story of abuse, memory, and contested truth.

The widely circulated photograph depicting Andrew Mountbatten-Windsor, Virginia Giuffre née Roberts, and Ghislaine Maxwell has been a focal point of public discourse for over a decade. From a forensic perspective, the case is highly unusual - not only due to the involvement of high-profile individuals, but also because of the unusual history of the image, which existed solely as a printed copy for a limited period. Further complicating matters is the notable absence of formal forensic analysis, despite widespread online speculation and commentary. Legal proceedings related to the case concluded in private settlement negotiations, making the image potentially one of the three most consequential photographs in modern history.

This study does not seek to influence public opinion, infringe on individual privacy, assign blame, offer legal advice or give final interpretation. The legal case has been resolved. The focus here is strictly on the photograph as an object of forensic inquiry by visual inspection and computational analysis. Nonetheless, the broader context - including the tragic consequences of sex trafficking - cannot be ignored. Jeffrey Epstein, a convicted sex offender, died in custody; Ghislaine Maxwell is serving a lengthy prison sentence. Andrew Mountbatten-Windsor has relinquished his royal titles. Despite these developments, many individuals associated with Epstein remain unprosecuted.

Tragically, several individuals identified as victims of Epstein's network, L. S. P. (d. 2017), C. A. (d. 2023), and most recently, Virginia Giuffre (d. 2025), have died (1). Their lives were irrevocably harmed (2). Ms. Giuffre’s allegations against Andrew Mountbatten-Windsor appear to have carried weight, given the resulting settlement. Yet central questions remain regarding the photograph itself: is it authentic, or a manipulated composite?

Public opinion remains divided. While most content creators argue for its authenticity, others have raised serious doubts (3)(4, 5). Hany Farid, an image expert from UC Berkeley, is cited as stating: “I can’t say for sure that the image is authentic, but I don’t see any obvious signs of manipulation.” Farid also noted that the light and shadows in the image show no inconsistencies, and it does not appear likely that Andrew’s body or head were spliced into the image (6). Joel van Hemel, another Florida-based legal image expert, "was unequivocally saying it was genuine."

This paper aims to contribute a first rigorous, scientific analysis to this debate, focusing on the technical characteristics of the image itself.

## Material and Methods

The analysis was initially drafted for the blog of the author, following the emergence of numerous questions regarding the photograph. However, it was not published at the time, as too many critical facts were missing. Following the death of Ms. Giuffre in 2025, the blog post was eventually developed into the present manuscript.

The current image analysis is based on publicly available versions of the photograph in question. The initial image found was the version released by the U.S. Department of Justice (7). Additional sources could also be examined, including the Daily Mail 2011 (DM2011, Fig. 1A) (8), 2023 (DM2023, Fig. 1B) (9) and most recently The Sun in 2019 (SUN2019, Fig. 1C) (10). As an anonymous X account (@RealMotherFaker  #281830) claimed access to a JPG file of dimensions 4896x3264 px representing the full sensor readout, the internet search was repeated; the highest resolution version found was SUN2019 at 4180x2858 px (10). Although it is also cropped compared to (@RealMotherFaker #281830), it appears to retain intact metadata.

An additional image of Ms. Giuffre, reportedly taken in 2001 and showing her in what appears to be the same outfit as in DM2011, was identified in coverage of an unrelated event in St. Tropez (4). Photographic archives such as Getty Images provide numerous contemporaneous images of Andrew Mountbatten-Windsor from around the year 2000 (11). With assistance from the Getty Images helpdesk and targeted searches of the PA Images archive, several closely related images of Andrew Mountbatten-Windsor were identified, yielding a shortlist of approximately ten candidates. Reverse-image searches using Google Image Search, PimEyes, and TinEye returned no additional photographs independent of the present context. Following submission of the initial manuscript, a further

email exchange (see Acknowledgments) produced an additional shortlist of Maxwell images.

Internet image searches were limited, however, as there are more than 11,000 copies of the image under scrutiny. For analysis of the candidate images eyes were detected in both images using Haar cascade classifiers. A two-dimensional transformation (rotation, scale, and translation) mapping the eye coordinates of the candidate image to those of the target image was estimated using estimateAffinePartial2D (OpenCV). The candidate image was then warped using warpAffine to achieve geometric alignment and to match the canvas dimensions of the target image.

To investigate image provenance, numerous tabloid articles were screened. Legal documents related to this case were downloaded (Epstein Estate Files, Giuffre v. Dershowitz, Giuffre v. Maxwell, Giuffre v. Prince Andrew) and indexed using an ad hoc Python script, which enabled natural language querying through a local Deepseek R1 0528 engine. Relevant forensic literature was also consulted, including foundational texts on digital image manipulation detection (12), (13). While some principles from these works were applicable, the specific characteristics of this case limited their full relevance.

To enable meaningful comparisons, independent photographic samples believed to originate from Kodak disposable and film cameras were also collected. Flickr hosts roughly 3,000 Kodak Funsaver images, and the Lomography archive another 2,000 images. Additionally, 20 further Kodak photographs taken by Giuffre from around the year 2000 were analyzed from the court files.

All images were processed using identical analytical methods. To characterize the noise and texture properties of each image and to compare them to reference samples of film-origin and digital-origin photographs, three frequency-domain descriptors were computed: (i) the power-spectral-density (PSD) slope $\alpha$ (14), (ii) the spectral flatness measure (SFM) (15) and (iii) the RGB-channel PSD correlation ρ_RGB (16). The PSD slope describes how rapidly spectral energy decays with frequency. Steeper, more negative slopes indicate film-like characteristics such as grain, whereas flatter slopes may reflect digital sensor noise.

Visual image analysis was performed using adjustment layers in Affinity Photo 1.10.8 as well as the forensic tool Sherloq (17), which includes principal component analysis (PCA) (18), (19) and luminance gradient analysis (LGA) (20), (21). A machine learning–based forensic technique was also employed. TruFor combines RGB features and a self-supervised noise fingerprint (Noiseprint++) via a transformer fusion architecture to detect image manipulations as deviations from expected camera-model patterns. (22).

Lastly, a full 3D reconstruction of the scene (23) was created using Blender 4.4, using fSpy for geometric camera alignment and floorplan registration. (23). These supplemental files can be accessed at https://github.com/under-score/little_prince.

ChatGPT 4.1 and 5.1 (OpenAI) with multimodal vision capabilities for direct image analysis yielded several notable observations, although not all of its conclusions were usable. ChatGPT was used also for programming of noise and texture parameter analysis but replaced by Anthropic's Claude Sonnet 4.6 following concerns raised in 2026 regarding OpenAI's conduct.

The ethical and legal questions surrounding the image are highly complex, as it depicts a scene on private property and involves two individuals who are not known to have consented. to its publication. As Ms. Giuffre had possessed the physical print, she has been compensated by a publisher for licensing a reproduction of the image even while not being the copyright holder that belongs to the photographer. Previous publications have carried contradictory copyright notices, including attributions to the U.S. District Court, DOJ, AFP, Getty Images, Picture Alliance, and Shutterstock. News outlets have published the image under fair use in the U.S. or fair dealing in the UK and other Commonwealth jurisdiction, which applies also to the present context as a publicly accessible, non-commercial scientific analysis of an image with no artistic value.

To preserve anonymity, names of actors not directly involved have been abbreviated.

## Results & Interpretation

Following contextual and origin-based analysis of the photograph, a more detailed visual examination was conducted including noise analysis, 3D scene reconstruction, and computational forensic analyses.

### *Historical Context*

Bloomberg News reported the opinion of Andrew Mountbatten-Windsor about the image (24):

"The Duke has previously said he has 'no recollection' of meeting Ms. Giuffre and cast doubt on the picture, claiming he is 'not one to hug' or 'display affection' in public." In an interview with BBC Newsnight in 2019, Andrew Mountbatten-Windsor said while he recognizes himself in the picture with Ms. Giuffre, he claimed it is not possible to prove whether the image had been faked. He also claimed the picture could not have been taken in London, as he was in his 'travelling clothes'. However, his explanation about attending a party at Pizza Express in Woking on that date was not widely accepted by the public (25).

Ghislaine Maxwell also questioned the image's authenticity during a deposition (26). She said "the surroundings in the photo looked 'familiar', but also said: 'We can't really establish the photograph and all that', adding: 'I don't know if that's true, if that's a real picture or not.' " This contrasts with an earlier email in which she seemed to  affirm the photo's authenticity (27).

Virginia Giuffre details the situation in her legal filing (7): "37. On one occasion, Prince Andrew sexually abused Plaintiff in London at Maxwell’s home. During this encounter, Epstein, Maxwell, and Prince Andrew forced Plaintiff, a child, to have sexual intercourse with Prince Andrew against her will. 38. The below photograph depicts Prince Andrew, Plaintiff, and Maxwell at Maxwell's home prior to Prince Andrew sexually abusing Plaintiff. [...] 41. During each of the aforementioned incidents, Plaintiff was compelled by express or implied threats by Epstein, Maxwell, and/or Prince Andrew to engage in sexual acts with Prince Andrew, and feared death or physical injury to herself or another and other repercussions for disobeying Epstein, Maxwell, and Prince Andrew due to their powerful connections, wealth, and authority. "

In addition there are numerous other accounts of the situation by Ms. Giuffre. These include her spiral notebook of roughly 300 pages, which she later reported that she had burned it (28); a 24-page set of diary-style notes prepared exclusively for Mail on Sunday journalist Sharon Churcher in 2011 (29) leading to the first Daily Mail article (9). There exists also an early, not dated autobiographical manuscript "The Billionaire's Playboy Club"( 30) and the posthumously published memoir "Nobody's Girl" by journalist Amy Wallace (31). In "Billionaire's Playboy Club" the photo is taken only after the group returns from Tramp nightclub; they chat briefly upstairs in the hallway, then Epstein snaps the picture. In "Nobody's Girl" Ms. Giuffre retrieves the disposable camera from her room upstairs and Epstein takes the photo in the entryway before going out for dinner. So timing and location vary substantially.

Recently surfaced private emails provide a more informal view and have been largely interpreted as confirming the authenticity of image. An email from Andrew Mountbatten-Windsor to Jeffrey Epstein on Monday, Feb 28, 2011, one day after initial publications reads, "I'm just as concerned for you! Don't worry about me!"(32). Another email from Epstein on July 1, 2011 states that "...the daily mail paid her money, they admitted it, ... Yes, she was on my plane, and yes she had her picture taken with Andrew, as many of my employees have. I have never misled you, this girl is a total liar..."(33). Andrew Mountbatten-Windsor's failure to protest immediately that the now-famous photograph was fabricated is noteworthy and may indicate that he understood it to refer to a real encounter with Ms. Giuffre. Epstein's own response is also unexpected. His wording could plausibly refer not to the disputed townhouse image but to another photograph taken during one of Andrew's documented flights on Epstein's jets or the "puppet" image taken in New York.

The presumed location is the second floor of Ghislaine Maxwell's residence in Belgravia, London (34). Later real estate listings confirmed that the image was taken on the first landing. The event was reportedly on Saturday, March 10, 2001, late at night, and according to Ms. Giuffre (9) was captured by Jeffrey Epstein using her personal camera, described to be her yellow Kodak FunSaver disposable camera. The film was allegedly developed at a Walmart photo lab near her former home in West Palm Beach, Florida, three days later.

The print reportedly remained among a dozen other images in a white envelope stored in a bookcase until journalist Sharon Churcher and photographer Michael Thomas visited Ms. Giuffre on February 17, 2011, in Glenning Valley, New South Wales after she filed a civil writ in Florida claiming she had been sexually exploited by friends of financier Epstein while under age. The photographer took numerous photos of the front and back side of the print but did not report having seen any negative or any additional photos from the London visit. Unfortunately, the reproduction lacks any size markers or color chart while an image of Ms. Giuffre holding the photo is also missing, although he took photos of her in at least five different outfits. A cropped version of his reproduction was then published in the first Daily Mail article on Sunday Feb 27, 2011 (updated on Mar 3, 2011 (8) alongside an interview, with Ms. Giuffre reportedly receiving $140,000 for the interview and the photo. The contract and all communication between Ms. Churcher and Ms. Giuffre before Feb 17, 2011, seems to have been lost.

The physical image was allegedly handed to the FBI at the U.S. embassy in Perth for reproduction before returned to her (35). The print however is also no longer in her possession and was reportedly lost afterwards (35). Following Ghislaine Maxwell's public denial of the image's authenticity, Michael Thomas provided further photographs to the Daily Mail on January 23, 2023 including also the photograph's reverse, suggesting it was printed on Kodak RA-4 paper, consistent with development on a roller-based minilab system. The back print reads "000 #15 13Mar01 Walgreens One Hour Photo," interpreted as: film or customer ID zero, 15th exposure, date, and lab name. According to press reports two witnesses, C.A. (now deceased) and T.F. (a former friend of Ms. Giuffre), confirmed having seen the image in 2001 (9).

Is there any possibility that the original photograph has been manipulated? If done between 2001 and 2011, both witnesses would need to have misremembered or knowingly provided false testimony. Forging a photograph to this extent would have required significant technical effort, though not impossible by 2010 standards. Considering the case's high profile and substantial financial settlements, and given that all parties involved have been inconsistent in past statements, such a possibility cannot be dismissed. Moreover, also journalistic sources may be unreliable as well (36). The Mail on Sunday has faced credibility concerns to the extent that it is even not considered as a reliable source on Wikipedia (37). Standard forensic protocols typically require a complete "audit trail" documenting the image's provenance from capture to publication - information that is unavailable in this case.

The publication trail raises also questions as DM2011, SUN2019 and DM2023 provide different information. When a newsroom receives a photo-of-a-photo (a print photographed under room light), the JPEG may look soft and flat, mid-tone contrast is low (paper reflection compresses gamma), and the color channels are correlated by warm indoor light. To make such an image printable, editors may have routinely boosted micro-contrast, giving this a more film-like appearance, which is not suspicious in itself; the apparent extra grain in DM2011 is not added for deception but for print legibility. According to the metadata records, the image was ingested a second time on Feb 1, 2015 into a different CMS instance, apparently this time by The Sun. Although Daily Mail and The Sun belong to different publishing groups (DMG Media and News UK, respectively), their printing and distribution operations have been consolidated under a joint operational structure the year before. On 13 August 2019 the file was then probably inadvertently exported from The Sun's WordPress system with intact metadata as SUN2019. The later, full-size DM2023 version was likely released as a rebuttal to suggestions that the image might have been fabricated - yet paradoxically introducing additional inconsistencies.

*General Remarks*

Although the image appears plausible at first inspection, discrepancies become evident when it is compared with photographs from the same camera provided by Ms. Giuffre during the Giuffre–Maxwell trial (Fig. 2A). Those reference images exhibit substantially lower spatial resolution than DM2023. Moreover, in contrast to unrelated photographs produced with the same camera model, the flash characteristics of DM2023 diverge from the expected optical properties (Fig. 2B). "In the hands of a talented forger, the methods for manipulating images may be applied so skillfully that the faked image appears authentic, even to a trained eye. But these same methods often leave anomalous patterns that are too regular to be accidental." (38).

Viewer psychology should also not be underestimated, as photographs convey implied knowledge about events - commonly described as the picture superiority effect (39). The image fulfills some voyeuristic expectations; its casual composition and use of a flash contribute to the appearance of amateur authenticity. It is a compositional effective image; Epstein is depicted as an inept photographer whose obstructing finger appears in the frame; Giuffre is presented in an idealized, almost luminously youthful manner; Andrew Mountbatten-Windsor displays a constrained, self-conscious smile suggestive of having been caught off-guard; and Maxwell appears positioned as the orchestrating presence in the background. The overall impression reads like an incidental snapshot which makes it understandable why the image elicited such rapid and widespread public resonance.

*Film vs. Digital Origin*

The DM2023 reproduction appears clean, intact, and free from any dog-earing in contrast what was previously described by Sharon Churcher. There is no evidence of fogging or X-ray artifacts, which were common in airport scans around the year 2000, particularly among frequent travelers.

The control images both from the online Kodak Funsaver collections and most of the Giuffre vs. Maxwell image collection behaved as expected from known photographic processes. They exhibited steep PSD slopes (−2.7 to −3.1), moderate flatness (0.6–0.7), modest RGB independence (0.6–0.8). These signatures point to a chain of film negative → scanner → RA-4 print, which matches the known operation of minilabs in the 2000s. By contrast, the SUN2019 image differs as its PSD slope was flatter ($\alpha \approx -1.98$), its spectral flatness was higher (≈0.81), and its RGB-channel correlation was high (>0.9). This constellation of values is not compatible with film-scan minilab prints.

A methodological concern arises from the fact that the SUN2019 image is itself a photograph of a physical print, raising the question of whether this re-capture process could overwrite or obscure the original noise characteristics. In general, a high-quality digital re-capture is expected to preserve the underlying film-grain PSD slope. An initial digital capture in 2001 is possible as digital compact cameras were entering mainstream consumer use around the turn of the millennium. If the primary source had been a digital image, subsequent manipulation could have been performed prior to printing. Any such operations would be permanently embedded (“baked”) into the printed output and could even visually mask irregularities in the underlying image.

*Print analysis*

It is assumed that the image was printed on Kodak RA-4 paper, based on backprint stamps typical of minilab outputs from that era (Fig. 3 ) (40). The date format DDMMMYY  is however atypical for an U.S.-based Walgreens lab and is more commonly associated with European or Australian lab output. Although previously cited as supporting evidence of authenticity (9), backprint stamps can be fabricated using secondhand matrix printers. Cut edges on both the front-right and rear-left sides do not appear smooth, as would be expected from automatic minilab cutters. When the reverse side of DM2023 is contrast-enhanced and vertically flipped, it reveals also a faint, translucent image of the front side. This is unexpected given that Kodak RA-4 paper is not typically translucent. and raises the possibility that two thin paper prints may have been glued together - one for the image and one for the backprint explaining the translucent bleed-through in Fig. 3. Without access to the physical print and without the ability to examine the cut edges for chemical traces, this conclusion remains speculative in particular as SUN2019 has no typical signs of an inkjet printer like chromatic dot patterns, and no feed-direction banding.

By the late 2000s, most 1-hour labs in drugstores/supermarkets were using digital minilabs, typically Fujifilm Frontier systems (e.g. Frontier 3x0, later 500-series) and Noritsu QSS series (QSS-30xx, 31xx, 34xx, 35xx, 37xx, etc.). These are “hybrid” machines: They can develop film (C-41 processor + film scanner) and they can print from digital files on CD, memory cards, USB, etc The exact pattern of the diagonal repeating “Kodak Paper” text with the box logo matches Kodak Edge backprints used between 1997 and mid-2000 which were in use still in 2010, especially in Australia, UK and Canada. The image could have been printed in the U.S but also in the UK as Boots is a UK subsidiary of Walgreens and uses the same type of one-hour minilab equipment. If a machine is configured to imprint the image’s meta date instead of the actual print date, it may also print ‘‘WALGREENS’ with the given date stamp, since the branding comes from shared system settings rather than from the country in which the print is produced. A technician in Australia could also re-program a minilab at any time, so the back print text and date do not necessarily reflect the country or the actual day of printing.

*Lighting and Shadow*

The flash behavior matches compact digital cameras far better than disposable film cameras. Early-2000s compact digital cameras employed higher-energy xenon flashes mounted close to the optical axis, yielding harder on-axis lighting. Disposable film cameras, equipped with weaker xenon tubes and wide reflectors, produced diffuse, rapidly decaying light, often leading to underexposure and grain-dominated shadow regions.

As the trigger of most cameras is top right, the bottom left white spot is difficult to explain. If the white spot were the photographer’s middle finger, it would sit only a few cm from the lens. Even with a small-format lens, depth of field is not infinite; a nearby finger would appear soft-edged, blurred, and gently shaded by the flash. Instead, the observed shape is sharp and uniformly white - optically impossible for a finger. At the bottom of the white region there is also a pointed triangular wedge. No part of a finger - tip, knuckle, or side - produces a hard-edged triangular apex, confirming this cannot be an anatomical intrusion.

The single-flash setup produces harsh highlights, catchlights in the eyes, and pronounced shadows. Notable inconsistencies exist: the legs of front subjects are unusually dark for their proximity to the flash, while the rear bannister ball appears brighter than the front, despite being slightly in shadow. Red-eye effects are present in the man, less so in the girl, and absent in the woman. Eye reflexes of Andrew Mountbatten-Windsor have a slight offset to the right. Shadows are missing or misplaced - for instance, the Virginia Giuffre casts no shadow under her chin, unlike Ghislaine Maxwell behind her. The rear shadow on the window frame in the next room is straight but should be rippled due to the uneven surface of the window frame.

*Image geometry and 3D model*

There is mild pincushion distortion of the image and horizontal moldings appear slightly misaligned. Geometric plausibility is largely maintained but exhibits some distortion. The window frame aligns nearly parallel with the film plane, enabling some geometric size estimation. The centered camera orientation results in direct flash reflection at the vanishing point, indicating the camera height was lower than expected for a 183 cm tall man and closer to the woman's eye level at ≈ 168 cm (Fig. 4).

There are inconsistencies in arm positioning of Mountbatten-Windsor and Giuffre. His left shoulder is unusually low while his arm seems wrapped around her ("gorilla arm"). Her right and left arm positions appear anatomically incongruent: the right shoulder is raised slighltly too high to pass beneath his armpit. If her right arm were behind his shoulder, her hand would not plausibly reach the side of his waist (Fig. 5). Moreover, there is no evident reason for her left hand not to rest on her hip (a common method to assess alleged anatomical inconsistencies is to physically reproduce the posture - typically in front of a mirror - to test its biomechanical plausibility).

The apparent spatial inconsistency can be further assessed through 3D scene reconstruction, as demonstrated in earlier studies (41), (38). The estimated height of Mountbatten-Windsor in the reconstructed 3D landing in Fig 4 is 173 cm, with an uncertainty range of 168–178 cm. This is notably below his actual height of ≈ 183 cm, considering that classic dress shoes can raise total stature to approximately 185–186 cm. In contrast, the estimated heights of both women appear consistent with their presumed height, supporting the accuracy of the reconstructed figure height.

*Artifacts*

There are many more inconsistencies and visual artifacts, which suggest further evidence of compositing (Fig. 6).

1. Wall leakage and shadow absence: The wall texture appears to bleed into the man's shirt, suggesting a segmentation error. No shadow is cast onto the wall by the subject, inconsistent with directional lighting.
2. Background line intrusion and hair reflection artifact: A dark background line overlaps the woman's shoulder. A light reflection in her hair forms an unnatural shape.
3. Open window and inconsistent temperature context: The window is open despite the reported exterior temperature of ~4°C at that evening in London. Subjects are dressed lightly for winter conditions.
4. Absence of perspiration contradicts prior dancing.
5. Tie artifact: A dark tie-like remnant at the waist may indicate a poorly removed accessory by image editing.
6. Shirt coloration appears between bodies where white wall should be visible.

7. Digit morphology anomaly: The fingers of the man are reddish; his index appears longer than the middle finger. There seems to appear also a sixth finger of the girl. The shadow is unplausible.

8. Unrealistic wall illumination and lack of shadow casting: The background wall is overlit despite foreground occlusion.

9. Edge discontinuity: A purse in the man's back pocket and an unnaturally sharp pocket edge may point to digital insertion.

10. Trousers lack texture.

These findings suggest potential compositing or some digital alteration beyond normal photographic processes.

*Computational forensics*

More advanced forensic methods are complicated here also by the fact that we are examining here a digital reproduction rather than a native digital file. Nevertheless, three analytical approaches could be employed. PCA is considered informative without any restriction. LGA analysis may provide new information under strong lighting conditions, whereas ELA is not considered reliable for the reproduction of a printed image.

PCA (Fig. 7A and 7B) is applied to detect statistical variance patterns. Some inconsistencies may indicate tampering as PCA accentuates structural edges and lighting transitions. Both Mountbatten-Windsor and Giuffre exhibit strong outlines, while Giuffre's hair and facial features appear more pronounced and more clearly separated from the body compared to the man. A halo effect is particularly visible around her head and shoulders; this suggests that her head and/or upper torso may have been edited separately from the rest of the image - commonly associated with head or body swaps. Mountbatten-Windsor's PCA edges are more cohesive, which may indicate less internal alteration or possibly a complete external insertion.

LGA (Fig. 8) is used to evaluate lighting consistency. Results could show implausible transitions between shaded and illuminated areas or implausible fall-off. Indeed, also LGA reveals unnatural edge behavior. The outlines of the people, especially around the woman's face, hair, chest, and right arm, display magenta fringes. These color separations are inconsistent with the smoother gradients observed in the surrounding environment. This may indicate abrupt changes in pixel structure, as one would expect from inserted elements. The arms, torsos, and backgrounds of the individuals are visually distinct in lighting direction and surface texture. As with PCA, the woman's head and torso exhibit discontinuous luminance transitions, supporting also the hypothesis of head or body swaps.

TruFor (22) produces a pixel-level integrity heatmap in which warmer colours indicate regions flagged as potentially manipulated and cooler colours indicate unaltered regions. In the output (Fig. 9), the figures return predominantly red and yellow tones, indicating elevated anomaly scores, while the right figure and background register in blue and cyan, consistent with an unmanipulated origin. The marked discontinuity at the boundary between the central and right figures is consistent with compositing from sources of differing provenance.

Taken together, the observer-independent forensic results align with the visual assessment that the individuals do not appear to originate from the same source.

*Candidate source images*

Of the approximately ten images initially shortlisted, one high-confidence source image of Andrew Mountbatten-Windsor has been identified to date. The image originates from a public appearance at the premiere of 102 Dalmatians on 6th

December 2000 at Leicester Square, London, covered by multiple photographers. The Getty Images archive contains a movie showing at least 30 flash exposures captured in rapid succession. The publicly available photograph (11) is not immediately identical to the image under analysis: minor differences in head proportions and coloration are apparent on first inspection (Fig. 10). Quantitative geometric comparison based on facial landmarks shows a geometric distance of 0.0032, indicating near-identity, while eye-based affine alignment yielded a reprojection error of 2.03 pixels - consistent with adjacent frames in a flash burst or modest manual geometric adjustment. The source image exhibits a slightly raised right eyebrow and minimally depressed upper lip, discrepancies consistent with normal facial micro-expressions and likely introduced during image preparation prior to compositing. The magnitude and character of the required adjustments are fully consistent with manual warp and liquify operations available in Adobe Photoshop CS4/CS5 (circa 2009–2010).

Three diagnostic features support the hypothesis that the candidate photograph served as the source image. In both images, the right eye is directed slightly to the left of the photographer, and a red-eye effect extends beyond the pupil into adjacent ocular regions, consistent with a manually applied effect. A strong rear light produces a characteristic edge flare along the neckline beneath the right ear, identical in both images and difficult to remove convincingly with tools available circa 2010. Finally, specular highlights from frontal flash illumination appear at the right cheek, central forehead, and right ear in both images; after alignment their relative positions and shapes remain consistent despite global differences in tone and color. Because such features depend on three-dimensional facial geometry, illumination geometry, and head orientation, their joint correspondence strongly supports a common photographic origin.

A source image for Ghislaine Maxwell has also been identified in a publicly documented photograph taken on 18 September 2000 at the Anand Jon Fashion

Show in New York, depicting M.K., D.T., and Ghislaine Maxwell (Fig. 11, photo by Patrick McMullan (42)). The temporal proximity to the Andrew source - both originating within the same three-month window of autumn 2000 - is consistent with a single coordinated image preparation effort. Quantitative forensic comparison yields a geometric distance of 0.003, marginally tighter than the Andrew result, while affine alignment produced a re-projection error of 2.52 pixels, within the range expected from pose and lighting variation between sessions. The elevated pixel residual of 50.76 reflects surface-level differences in lighting and color grading after geometric alignment, not structural divergences.

Specular highlights appear at consistent facial locations in both images - central forehead, nose bridge, right cheek, and upper lip - and after geometric alignment their positions and shapes remain stable despite global differences in tone and brightness. Because highlight placement depends simultaneously on three-dimensional surface geometry, illumination angle, and head orientation, their correspondence across two ostensibly independent photographs supports a common photographic origin. The hair contour shows visible modification relative to the source: the McMullan photograph depicts a fuller crown and softer hairline, whereas the analyzed image shows a sharper hairline and slightly reduced crown, consistent with manual masking and hair replacement during compositing. The white collar or wrap visible at the neckline is geometrically consistent across both images after alignment. Given the difficulty of convincingly reconstructing fabric folds and cloth highlights from scratch, this correspondence further supports shared photographic origin rather than independent capture.

Discussion

This analysis set out to examine a widely circulated image with a neutral, forensic approach, independent of legal, political, or emotional narratives. While the absence of an original negative or uncompressed digital source limits the scope of definitive conclusions, key reproductions offered a sufficient foundation for a technical review. A range of visual inconsistencies was identified, spanning lighting anomalies, anatomical proportions and contextual contradictions. While some of these may result from benign photographic conditions, others suggest digital or physical alteration.

It is essential to reiterate that a scientifically robust forensic evaluation typically requires access to the original medium - whether film negative or native digital file - with a complete and documented audit trail. In this case, provenance assessment is based on journalistic descriptions of a mysteriously vanished print that was digitized under unknown conditions and even edited later. This inherently introduces interpretive uncertainty. Nevertheless, specific indicators - such as implausible lighting gradients, edge discontinuities, and scale distortion - mirror known inconsistencies in composite imagery.

Another dimension of concern lies in the image's production, handling, and release by the publisher. The 2023 reproduction set includes new photographs of the print, yet it is unclear why so many exposures were necessary and even none could be provided here for analysis. The fact that the full image reproduction was not released until over a decade after its alleged discovery raises legitimate questions.

There is also the unresolved issue of the missing print. Given the high-profile nature of the individuals depicted, the absence of such a fundamental component is striking. Why would a victim of sex trafficking retain what might be viewed as a "trophy image" taken for her family? Ms. Giuffre is cited as saying "I wanted to show something to my mom" (1). And why would this image then be shared with a journalist - accompanied by an interview and high financial compensation - yet no effort made to secure or preserve the rest of the roll? Twelve or more images would have likely followed this exposure on the putative 27-frame roll, yet none have surfaced. These gaps in the narrative undermine the ability to place the image within a consistent temporal context and suggest either highly unusual storage conditions or intentional curation.

Both soft and hard evidence suggest that this image may be a digital composite. Hard evidence includes the digital origin, the unnatural body height of Andrew Mountbatten-Windsor and the misplaced shadow near Ms. Giuffre. There are many more subtle but convincing anatomical inconsistencies of Ms. Giuffre's and Andrew Mountbatten-Windsor's arms. While no single irregularity offers definitive proof, the convergence of multiple indicators strengthens the case for image manipulation.

However, the specifics - when, how, and by whom - remain entirely unknown. Until additional evidence emerges, such as an original negative or positive, more related photographs or the files showing the manipulation, this mystery remains unresolved. Absent such disclosures, direct testimony from someone with firsthand knowledge could also fill the remaining gaps. One plausible scenario at the moment - based on the analysis presented here - is that the original image was taken by Ms. Giuffre herself. In such a case, Ms. Giuffre may have had additional contemporaneous photographs of herself available and of course the public image of Andrew Mountbatten-Windsor and Ghislaine Maxwell. Whether any such work was carried out with assistance from friends, collaborators, or media professionals remains unknown, and warrants further investigation that is far beyond the scope of the analysis presented here.

The findings presented here neither exonerate Andrew Mountbatten-Windsor nor confirm Ms. Giuffre's account of the occasion depicted. In light of her later testimony in the Dershowitz case, the inconsistencies noted in her account of a family matter and the traffic incident before her death, the possibility of inaccuracies in her account cannot be excluded, particularly given that polygraph testing was administered by her own lawyer, David Boies. Memory inconsistencies have been reported in individuals with a history of prior trauma (43).

The question of why such manipulation would occur warrants consideration. What motive would justify altering a photograph of this nature? As Farid notes, Stalin, Mao, Hitler, Mussolini, Castro, Brezhnev, and many others had photographs manipulated in an attempt to rewrite history, as they understood that changing the visual record could change history itself (38). In this case, it is conceivable that a victim of abuse used an image to shift power dynamics - to reverse the narrative, or at least control a part of it. The image may represent a form of psychological self-defense - a way to reclaim dignity by visually framing a moment that preceded trauma. Alternatively, it could be interpreted as a retaliatory gesture for perceived injustice but also simply a calculated effort to secure financial compensation. As Palmeri observed in Politico magazine, in Epstein's world women both were victims of a hostile environment and sometimes also reaped the benefits of their association with him, or worse (2). Whatever the underlying motivation, the image appears to have achieved a profound impact on public perception, suggesting that it fully accomplished the purpose for which it was created - whatever that purpose may have been.

In any case, this contributes to the long history of the British Crown's association with doctored photographs. It is not limited to the image examined here; it extends to the 2024 Princess Catherine Mother's Day photograph ("Kategate") or the Balmoral family portrait of Queen Elizabeth II, which featured mismatched tartan patterns and cloned hair. Royal photographer Cecil Beaton had already extensively retouched royal images in the 1940s-1950s, slimming waistlines and removing double chins and wrinkles throughout his official work (44). The only difference now is that image manipulation appears not to have been carried out for favorable presentation, but rather to inflict reputational harm.

Ultimately, the photograph presented to photographer Michael Thomas in 2011 must be viewed as a single, uncertain fragment of a much larger and deeply tragic narrative. It offers, at most, a partial view into a complex story of abuse, power, memory, and media - but is not a final word on what truly happened.

Fig. 1A: DM2011 is the first published image, a grainy cropped image. Extracted metadata is given next to the image.

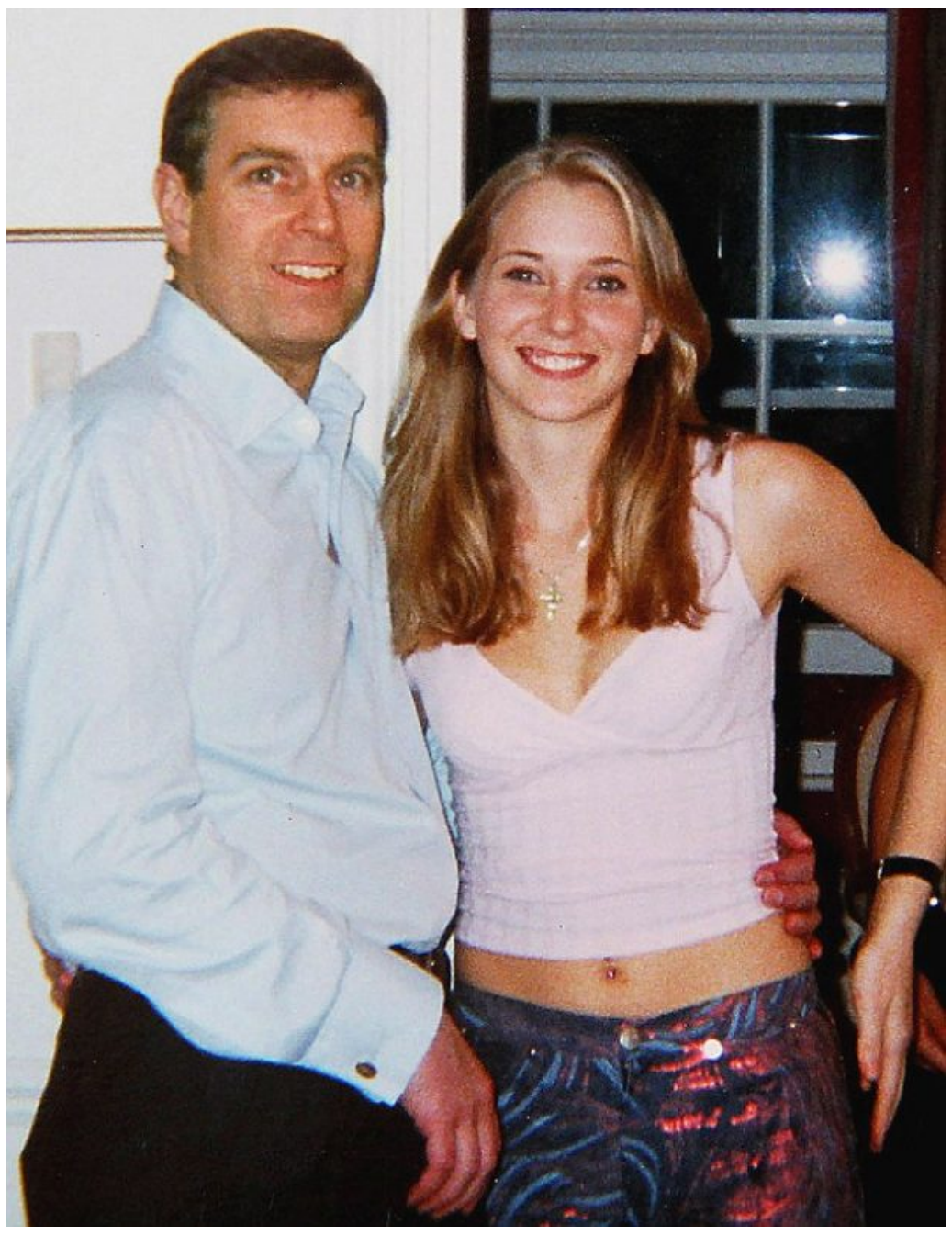

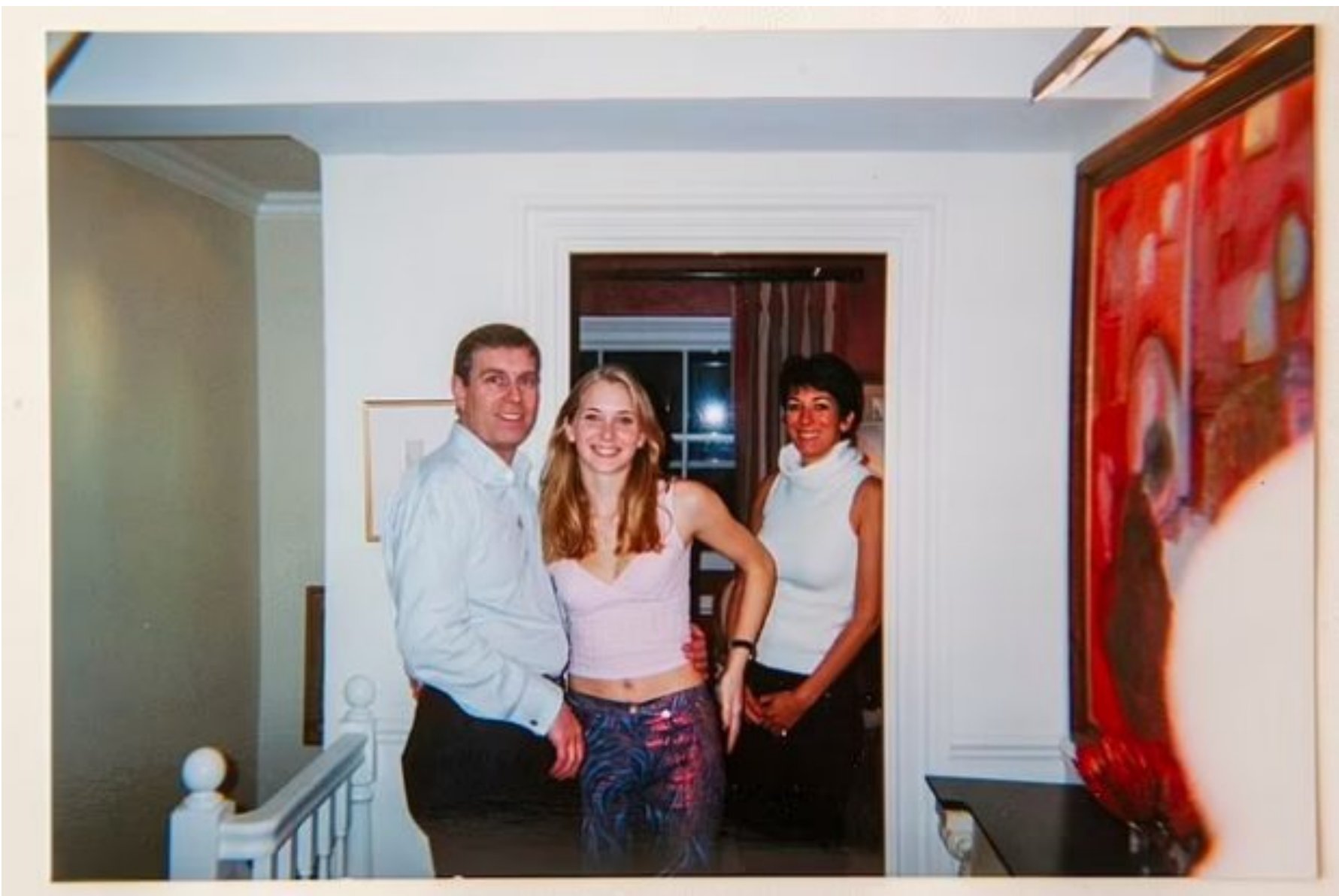

Fig. 1B: DM2023 uses the same source as DM2011 but is showing the full print.

Fig. 1C: SUN2019 displays the highest resolution and detailed metadata but is also cropped when compared to DM2023.

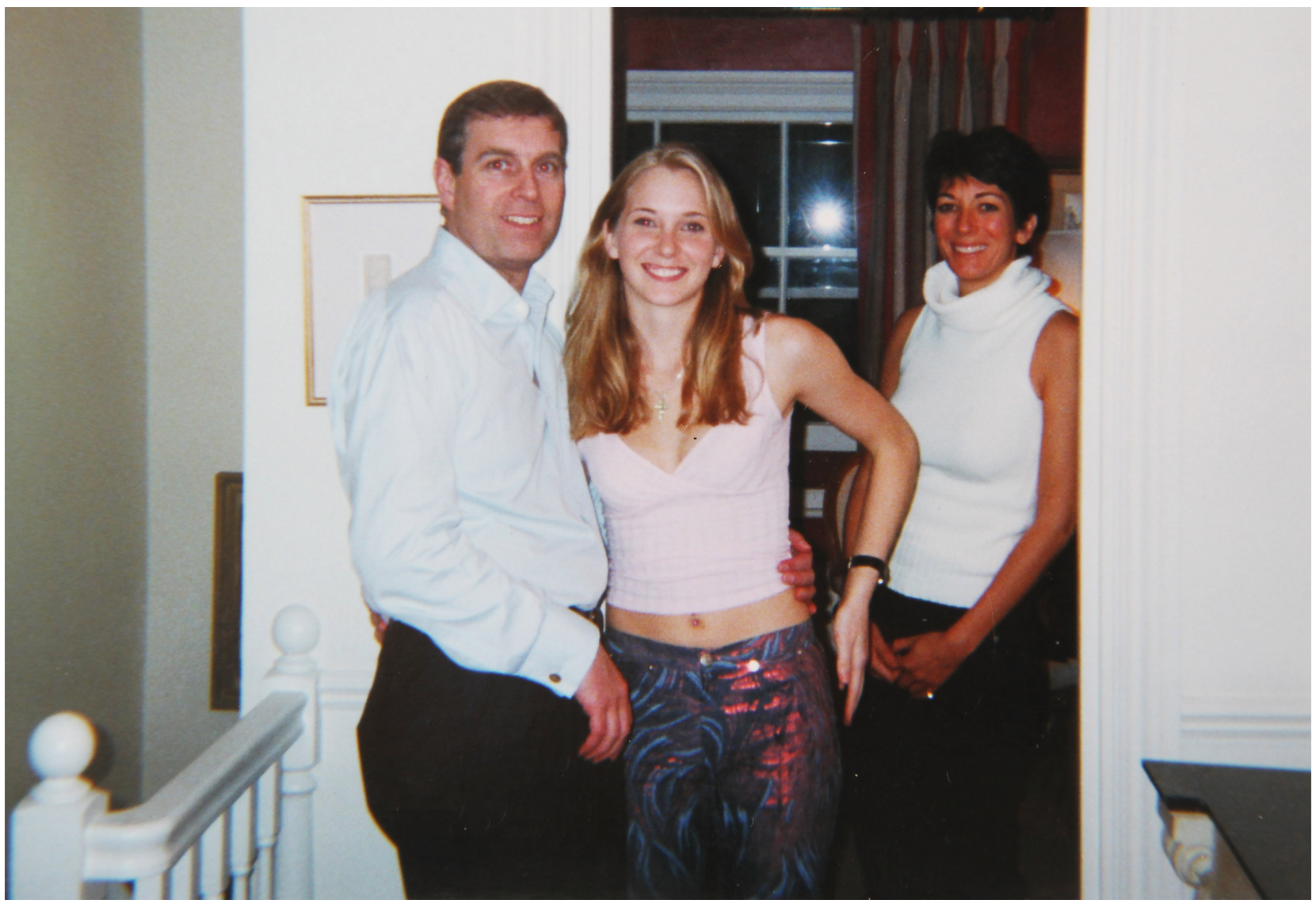

```
Image Description               : Virginia Roberts photographed with Andrew Mountbatten-
Windsor and Ghislaine Maxwell in early 2001. Collect picture from Virginia Roberts.
Make                            : Canon
Camera Model Name               : Canon EOS-1D Mark IV
Orientation                     : Horizontal (normal)
X Resolution                    : 300
Y Resolution                    : 300
Resolution Unit                 : inches
Software                        : Adobe Photoshop CS4 Macintosh
Modify Date                     : 2011:02:17 22:42:22
Y Cb Cr Positioning             : Co-sited
Exposure Time                   : 1/400
F Number                        : 8.0
Exposure Program                : Manual
ISO                             : 320
Exif Version                    : 0221
Date/Time Original              : 2011:02:17 13:04:27
Create Date                     : 2011:02:17 13:04:27
Components Configuration        : Y, Cb, Cr, -
Shutter Speed Value             : 1/395
Aperture Value                  : 8.0
Exposure Compensation           : 0
Metering Mode                   : Multi-segment
Flash                           : Off, Did not fire
Focal Length                    : 105.0 mm
User Comment                    :
Sub Sec Time                    : 75
Sub Sec Time Original           : 75
Sub Sec Time Digitized          : 75
Flashpix Version                : 0100
Color Space                     : sRGB
Exif Image Width                : 4180
Exif Image Height               : 2858
Interoperability Index          : R98 - DCF basic file (sRGB)
Interoperability Version        : 0100
Focal Plane X Resolution        : 3795.348837
Focal Plane Y Resolution        : 3904.30622
```

```
Focal Plane Resolution Unit     : inches
Custom Rendered                 : Normal
Exposure Mode                   : Manual
White Balance                   : Auto
Scene Capture Type              : Standard
GPS Version ID                  : 2.2.0.0
Compression                     : JPEG (old-style)
Thumbnail Offset                : 1288
Thumbnail Length                : 5036
Current IPTC Digest             : c4a40ad40a1ff6f352d452949eba49ff
Object Name                     : NINTCHDBPICT000141368077
Fixture Identifier              : 141368077
Keywords                        : royal family, Virginia Roberts, Andrew Mountbatten-
Windsor, Ghislaine maxwell
Special Instructions            :
By-line                         :
By-line Title                   :
Sub-location                    :
Province-State                  :
Country-Primary Location Name   : Australia
Original Transmission Reference : NINTCHDBPICT000141368077
Copyright Notice                : Collect
Caption-Abstract                : ***MUST PIXELLATE FACE OF VIRGINIA ROBERTS (CENTRE)***
***PLEASE CHECK WITH PICTURE EDITOR BEFORE PUBLICATION ANYWHERE*** 17 year old Virginia
Roberts photographed with Andrew Mountbatten-Windsor in early 2001. Collect picture from
Virginia Roberts. Roberts has accused billionaire businessman and convicted paedophile
Jeffrey Epstein of sexual exploitation.
Writer-Editor                   :
Displayed Units X               : inches
Displayed Units Y               : inches
Print Style                     : Centered
Print Position                  : 0 0
Print Scale                     : 1
Global Angle                    : 30
Global Altitude                 : 30
URL List                        :
Slices Group Name               : MJT_7889
Num Slices                      : 1
Pixel Aspect Ratio              : 1
Photoshop Thumbnail             : (Binary data 5036 bytes, use -b option to extract)
Has Real Merged Data            : Yes
Writer Name                     : Adobe Photoshop
Reader Name                     : Adobe Photoshop CS4
Photoshop Quality               : 9
Photoshop Format                : Standard
XMP Toolkit                     : XMP Core 5.1.2
NI Prepress Master ID           : 169759848
Caption Writer                  :
Headline                        : Billionaire Businessman and Convicted Paedophile
Jeffrey Epstein
Authors Position                :
Credit                          :
City                            :
Country                         : Australia
Transmission Reference          : NINTCHDBPICT000141368077
Date Created                    : 2015:01:02
State                           :
Instructions                    :
Source                          : Collect
Description                     : ***MUST PIXELLATE FACE OF VIRGINIA ROBERTS (CENTRE)***
***PLEASE CHECK WITH PICTURE EDITOR BEFORE PUBLICATION ANYWHERE*** 17 year old Virginia
Roberts photographed with Andrew Mountbatten-Windsor in early 2001. Collect picture from
Virginia Roberts. Roberts has accused billionaire businessman and convicted paedophile
Jeffrey Epstein of sexual exploitation.
Title                           : NINTCHDBPICT000141368077
Subject                         : royal family, Virginia Roberts, Andrew Mountbatten-
Windsor, Ghislaine maxwell
Rights                          : Collect
Location                        :
Metadata Date                   : 2019:08:12 21:58:45+01:00
Profile CMM Type                : Linotronic
Profile Version                 : 2.1.0
Profile Class                   : Display Device Profile
Color Space Data                : RGB
```

```
Profile Connection Space        : XYZ
Profile Date Time               : 1998:02:09 06:49:00
Profile File Signature          : acsp
Primary Platform                : Microsoft Corporation
CMM Flags                       : Not Embedded, Independent
Device Manufacturer             : Hewlett-Packard
Device Model                    : sRGB
Device Attributes               : Reflective, Glossy, Positive, Color
Rendering Intent                : Media-Relative Colorimetric
Connection Space Illuminant     : 0.9642 1 0.82491
Profile Creator                 : Hewlett-Packard
Profile ID                      : 0
Profile Copyright               : Copyright (c) 1998 Hewlett-Packard Company
Profile Description             : sRGB IEC61966-2.1
Media White Point               : 0.95045 1 1.08905
Media Black Point               : 0 0 0
Red Matrix Column               : 0.43607 0.22249 0.01392
Green Matrix Column             : 0.38515 0.71687 0.09708
Blue Matrix Column              : 0.14307 0.06061 0.7141
Device Mfg Desc                 : IEC http://www.iec.ch
Device Model Desc               : IEC 61966-2.1 Default RGB colour space - sRGB
Viewing Cond Desc               : Reference Viewing Condition in IEC61966-2.1
Viewing Cond Illuminant         : 19.6445 20.3718 16.8089
Viewing Cond Surround           : 3.92889 4.07439 3.36179
Viewing Cond Illuminant Type    : D50
Luminance                       : 76.03647 80 87.12462
Measurement Observer            : CIE 1931
Measurement Backing             : 0 0 0
Measurement Geometry            : Unknown
Measurement Flare               : 0.999%
Measurement Illuminant          : D65
Technology                      : Cathode Ray Tube Display
Red Tone Reproduction Curve     : (Binary data 2060 bytes, use -b option to extract)
Green Tone Reproduction Curve   : (Binary data 2060 bytes, use -b option to extract)
Blue Tone Reproduction Curve    : (Binary data 2060 bytes, use -b option to extract)
DCT Encode Version              : 100
APP14 Flags 0                   : [14]
APP14 Flags 1                   : (none)
Color Transform                 : YCbCr
Image Width                     : 4180
Image Height                    : 2858
Encoding Process                : Baseline DCT, Huffman coding
Bits Per Sample                 : 8
Color Components                : 3
Y Cb Cr Sub Sampling            : YCbCr4:4:4 (1 1)
Aperture                        : 8.0
Image Size                      : 4180x2858
Megapixels                      : 11.9
Scale Factor To 35 mm Equivalent: 1.1
Shutter Speed                   : 1/400
Create Date                     : 2011:02:17 13:04:27.75
Date/Time Original              : 2011:02:17 13:04:27.75
Modify Date                     : 2011:02:17 22:42:22.75
Thumbnail Image                 : (Binary data 5036 bytes, use -b option to extract)
Circle Of Confusion             : 0.027 mm
Field Of View                   : 17.6 deg
Focal Length                    : 105.0 mm (35 mm equivalent: 116.4 mm)
Hyperfocal Distance             : 50.83 m
Light Value                     : 13.0
```

Fig. 2A: Other images (in question also from Kodak Funsaver) released during the Giuffre vs. Maxwell trial. Despite the limited image quality yet the London townhouse image displays notably brighter scene illumination.

Fig. 2B: Unrelated random flash photographs tagged as Kodak FunSaver on Flickr that may illustrate the camera's characteristic light fall-off. The top-right image shows partial flash obstruction, while the bottom-left and middle-right examples exhibit true lens obstruction by a finger.

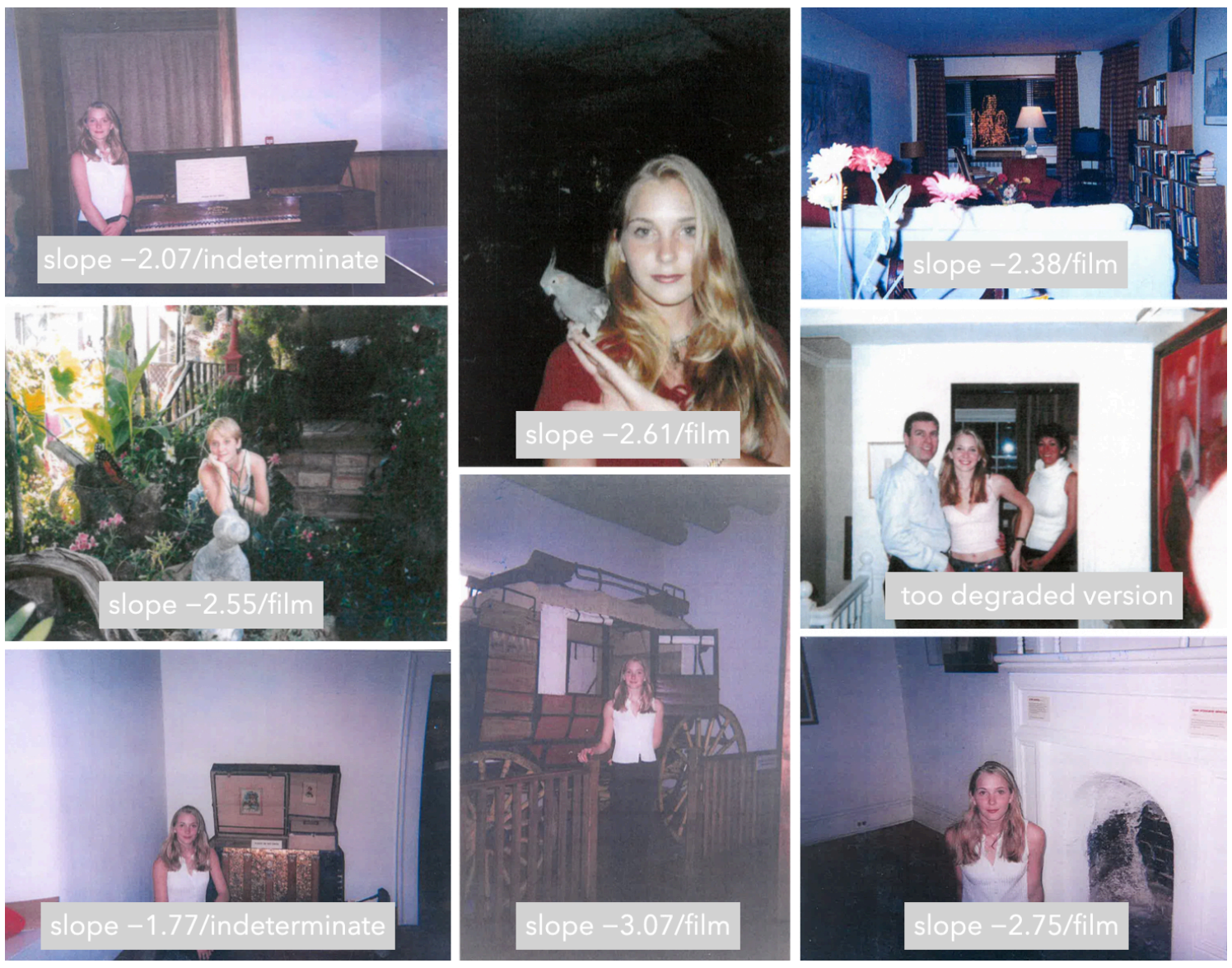

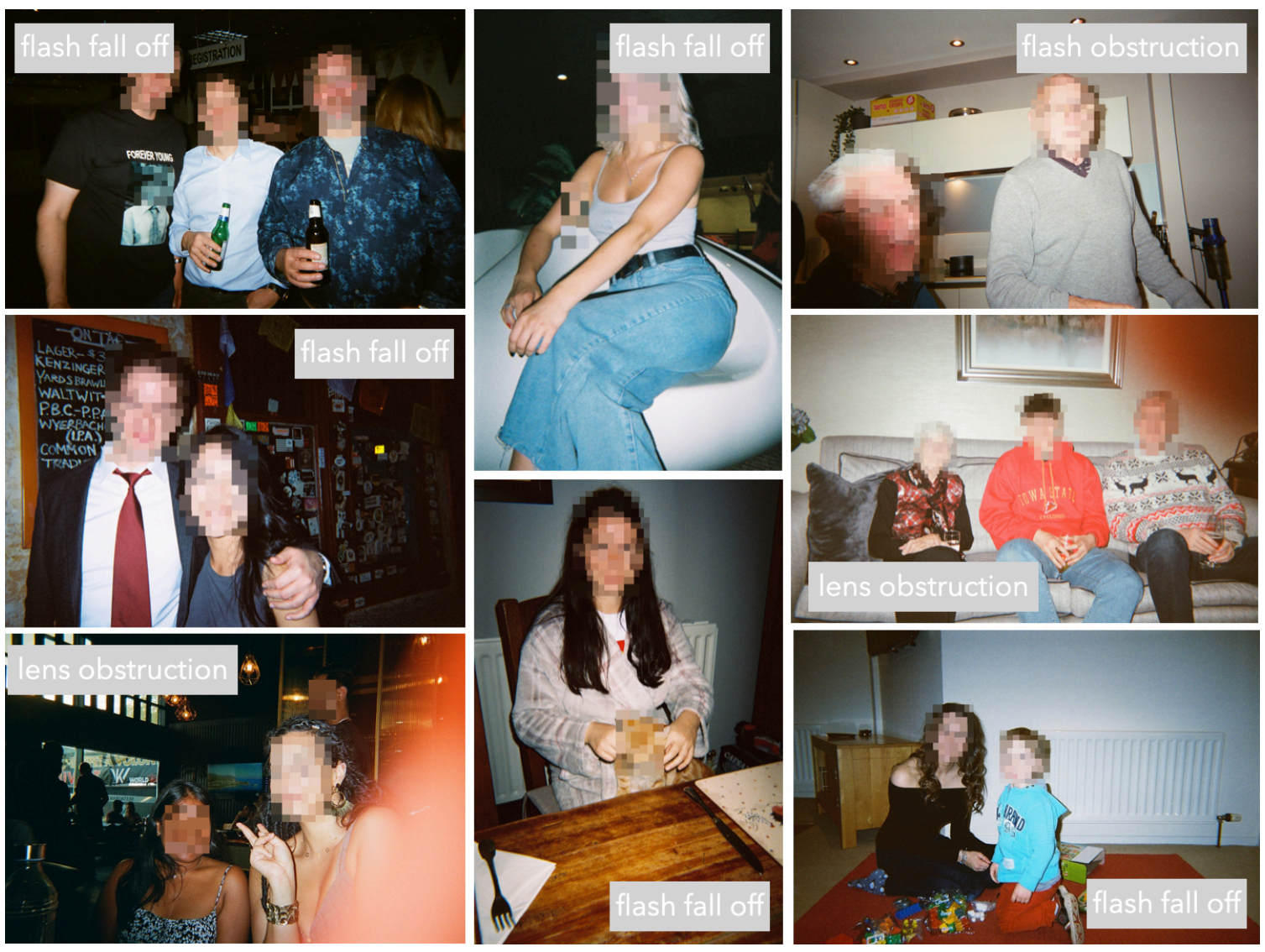

Fig. 3A: The back side of DM2023.

Fig. 3B: The reverse side of DM2023 is contrast-enhanced and vertically flipped. To restore the correct orientation, a horizontal flip is also required. This operation returns the text to its readable (non-mirror-reversed) form and allows the front image to shine through.

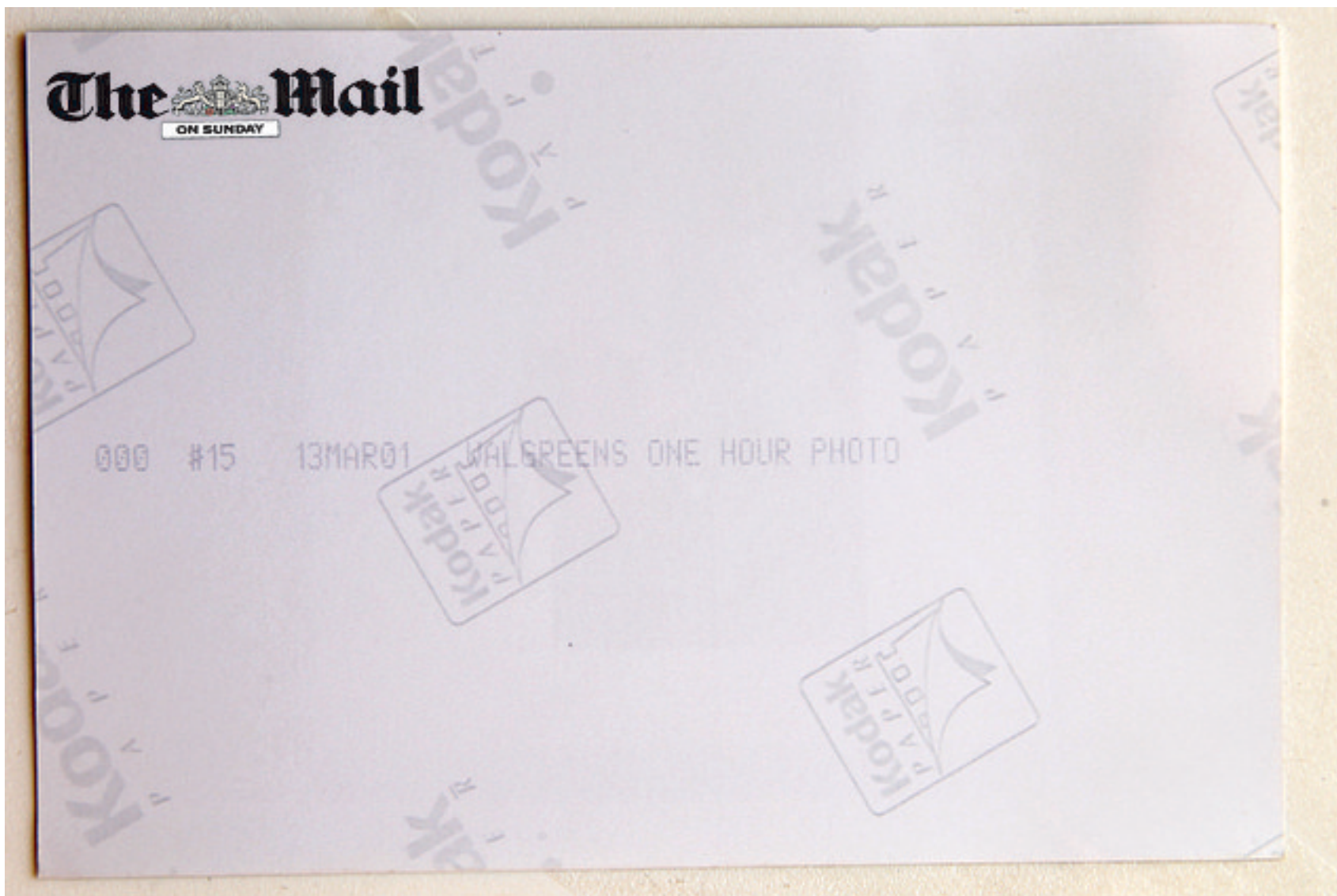

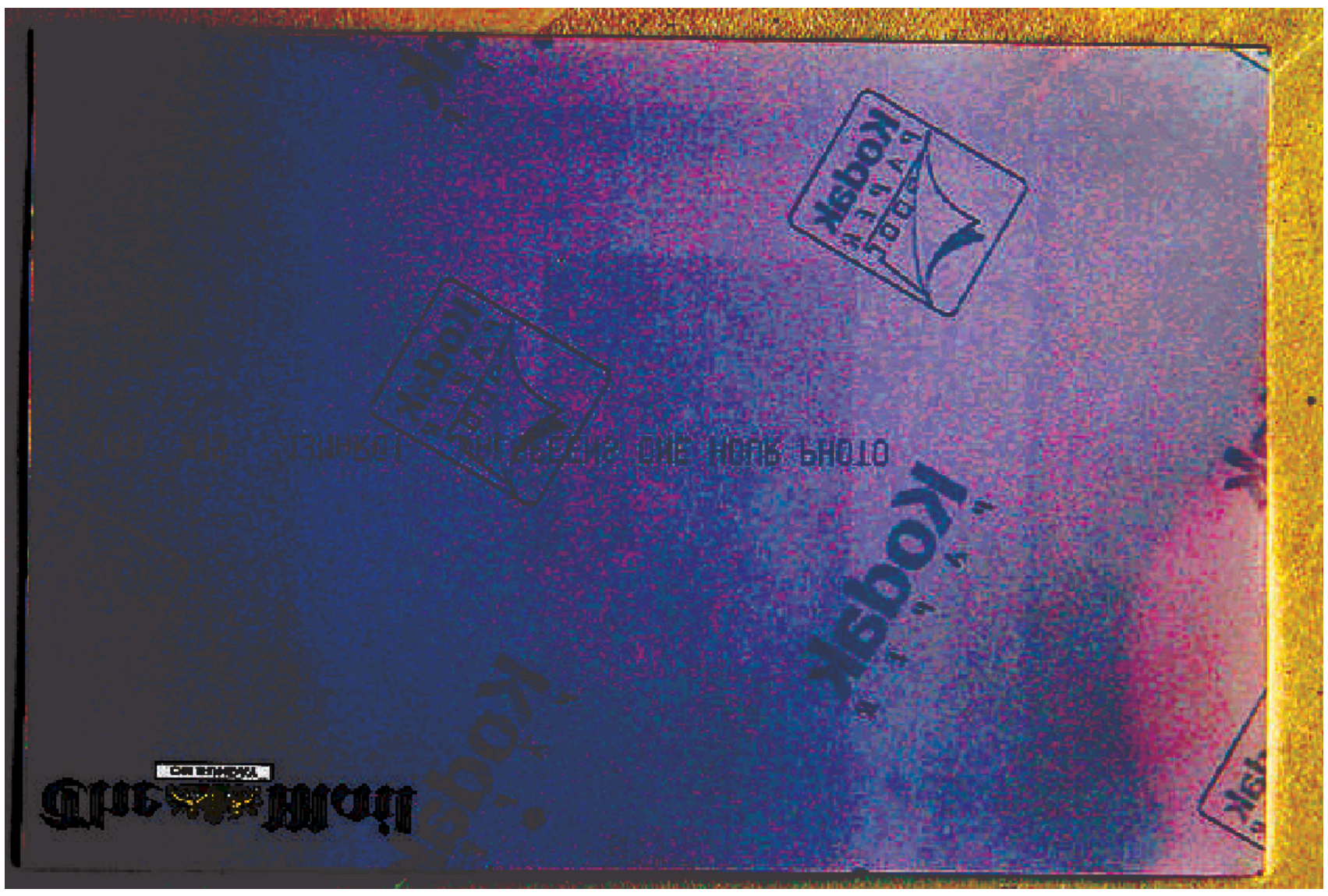

Fig. 4: Vanishing point  reconstruction of DM2023 using FSpy. It is situated t the center of the flash reflection in the rear window indicating a camera position at eye level of the woman.

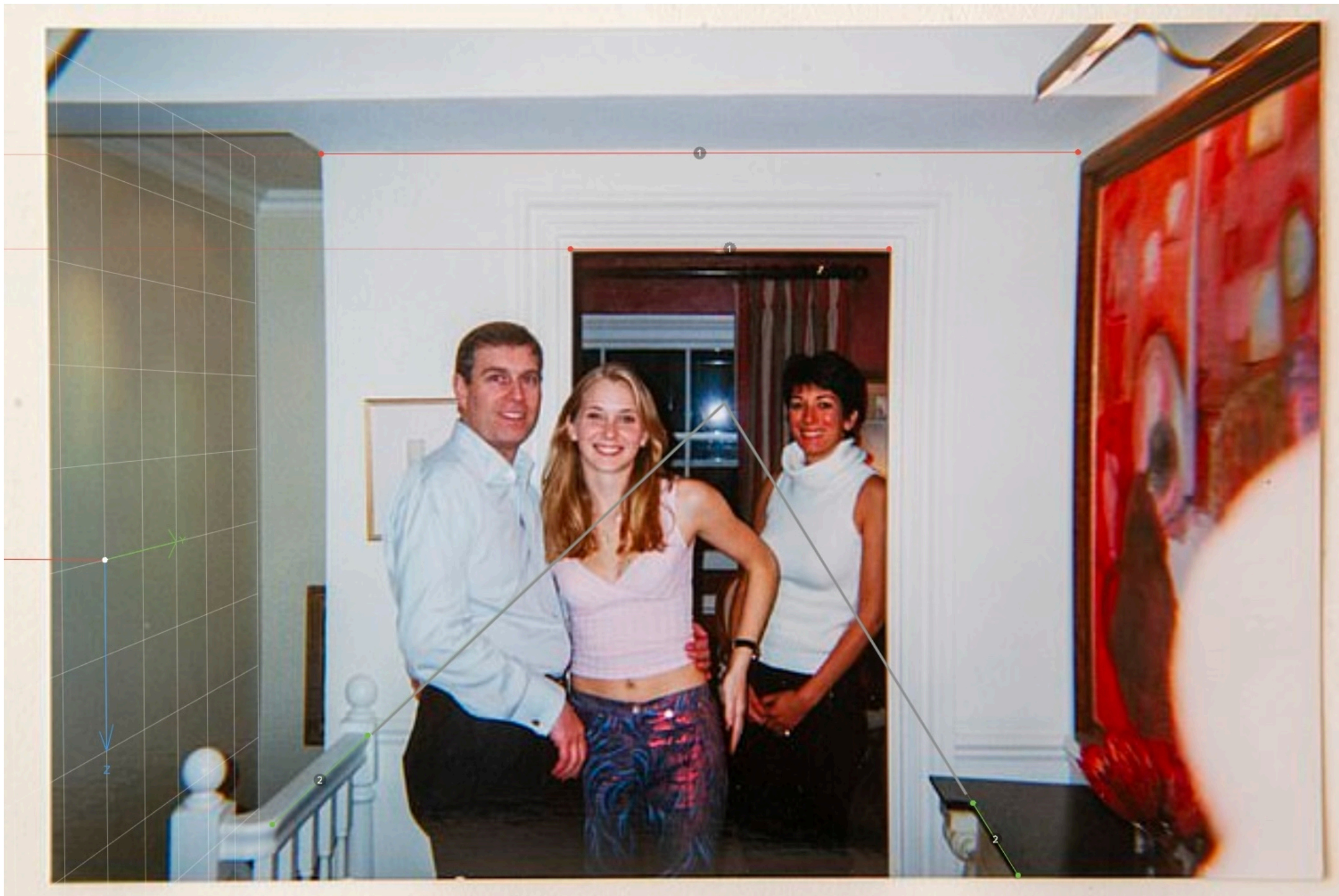

Fig. 5: A three-dimensional reconstruction of the first floor landing using DM2023 geometry and dimensions of the published floor plan. Andrew Mountbatten-Windsor's height in the photograph is estimated to be 173 cm although it should be >183 cm. The right arm of Virginia Giuffre has an anatomically impossible position, intersecting the lateral chest of Andrew Mountbatten-Windsor. Her left arm is supposed to rest on the hip but it is free-floating.

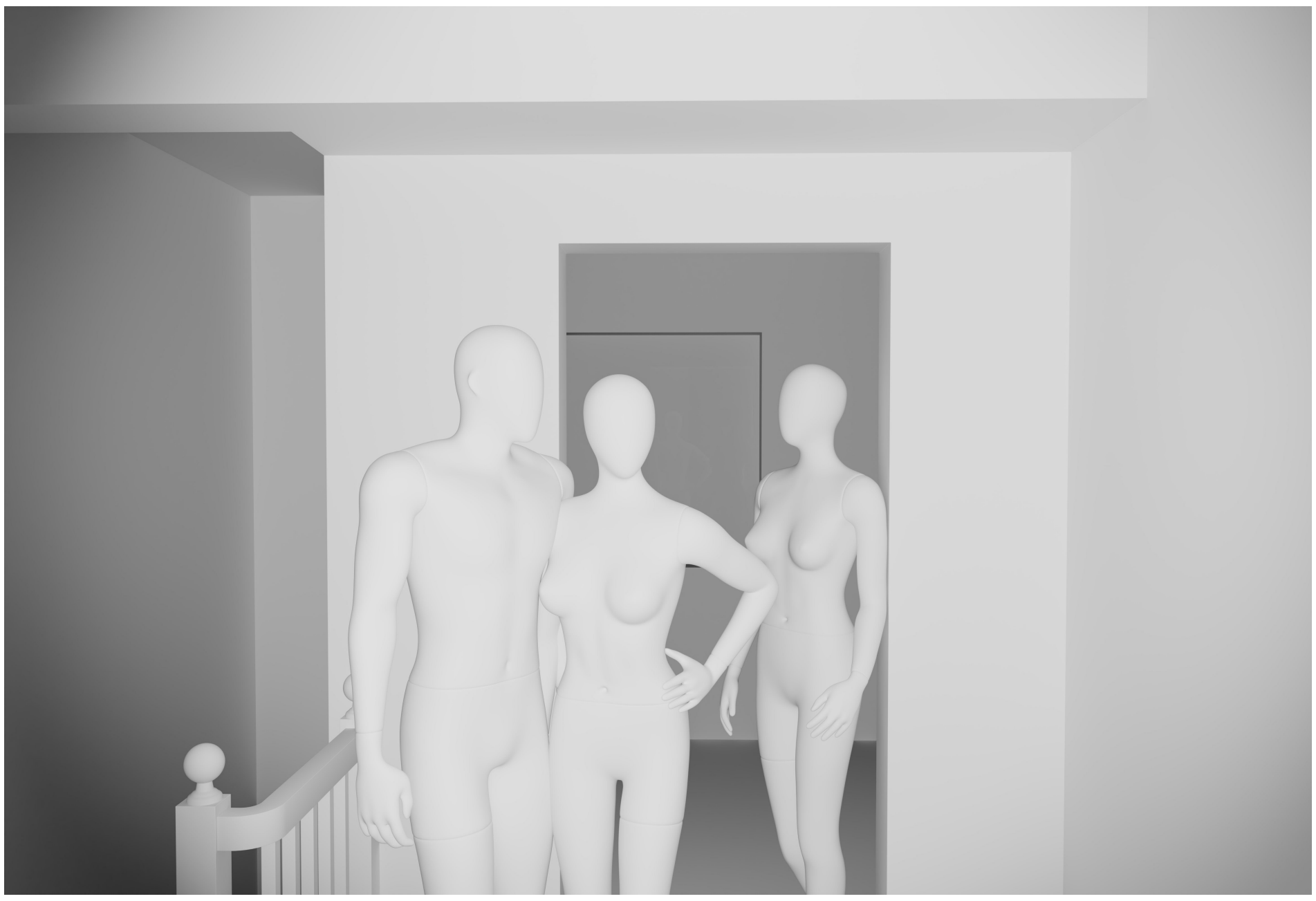

Fig. 6: Visual irregularities found in DM2011. See text for legends.

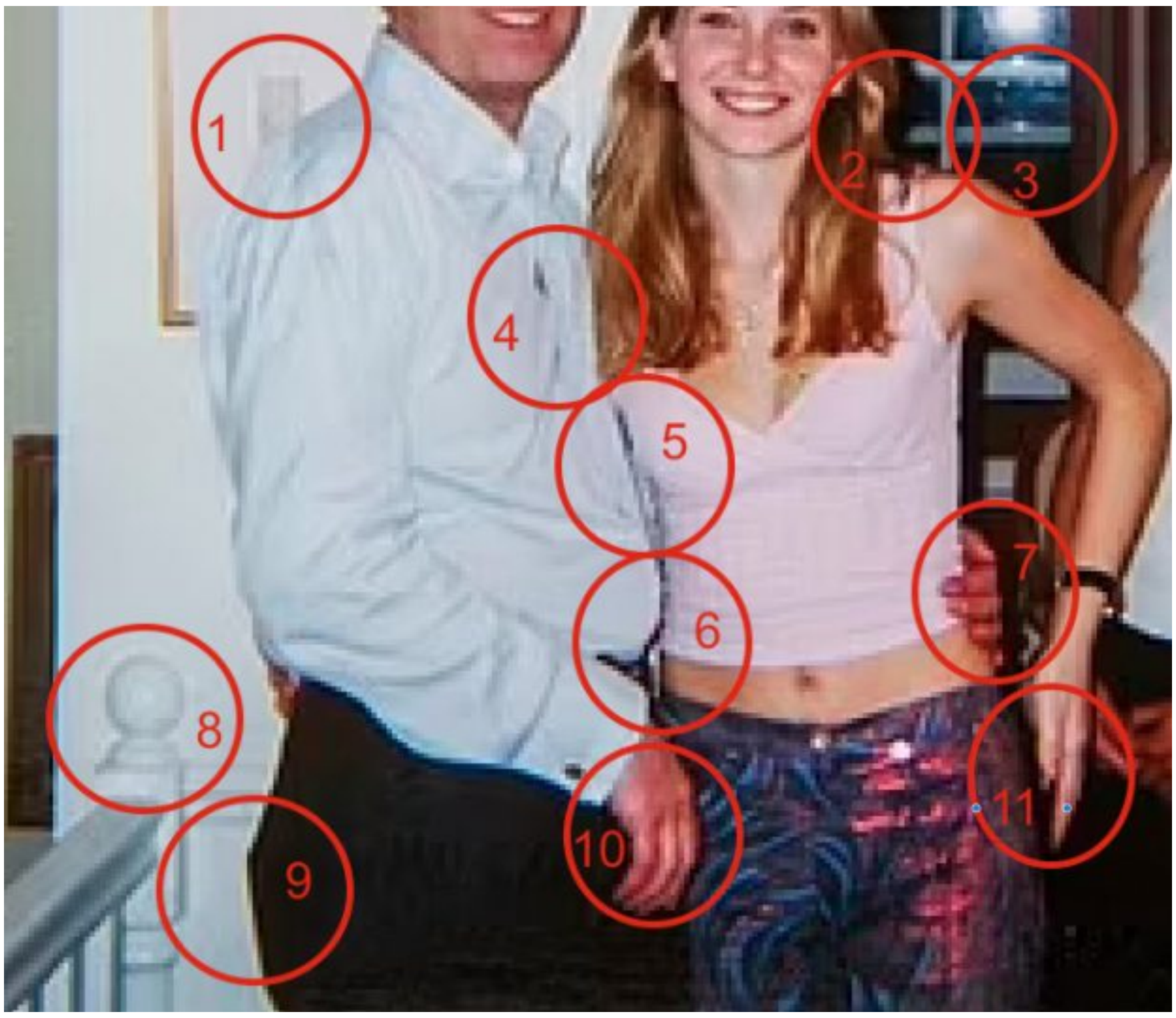

Fig. 7: Principal Component Analysis (PCA) of DM2011 distances. First component is shown at the left and second component is shown inverted at the right.

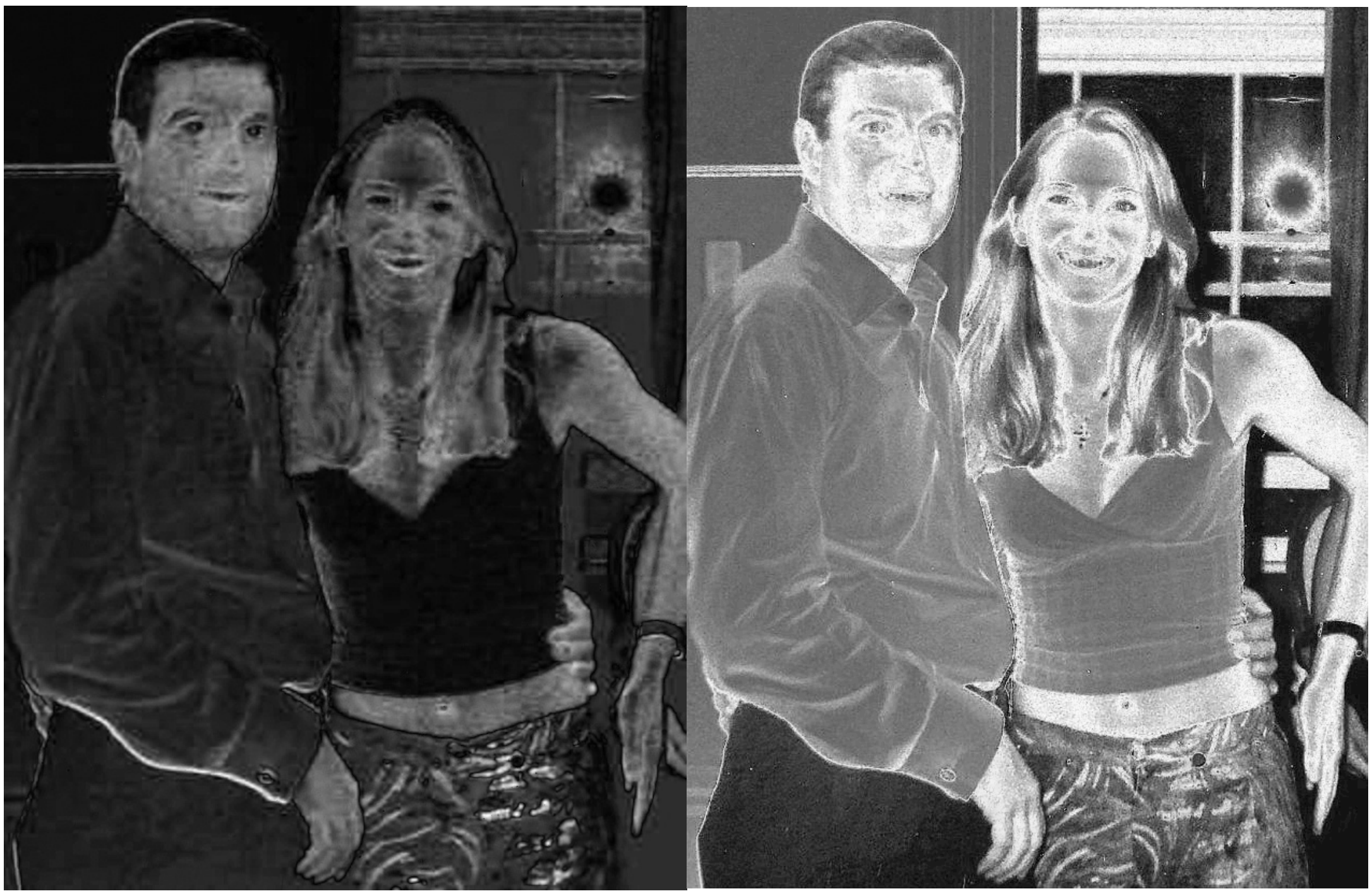

Fig. 8: Luminance gradient analysis LGA of DM2011 with settings 95% intensity, blue channel, normalized.

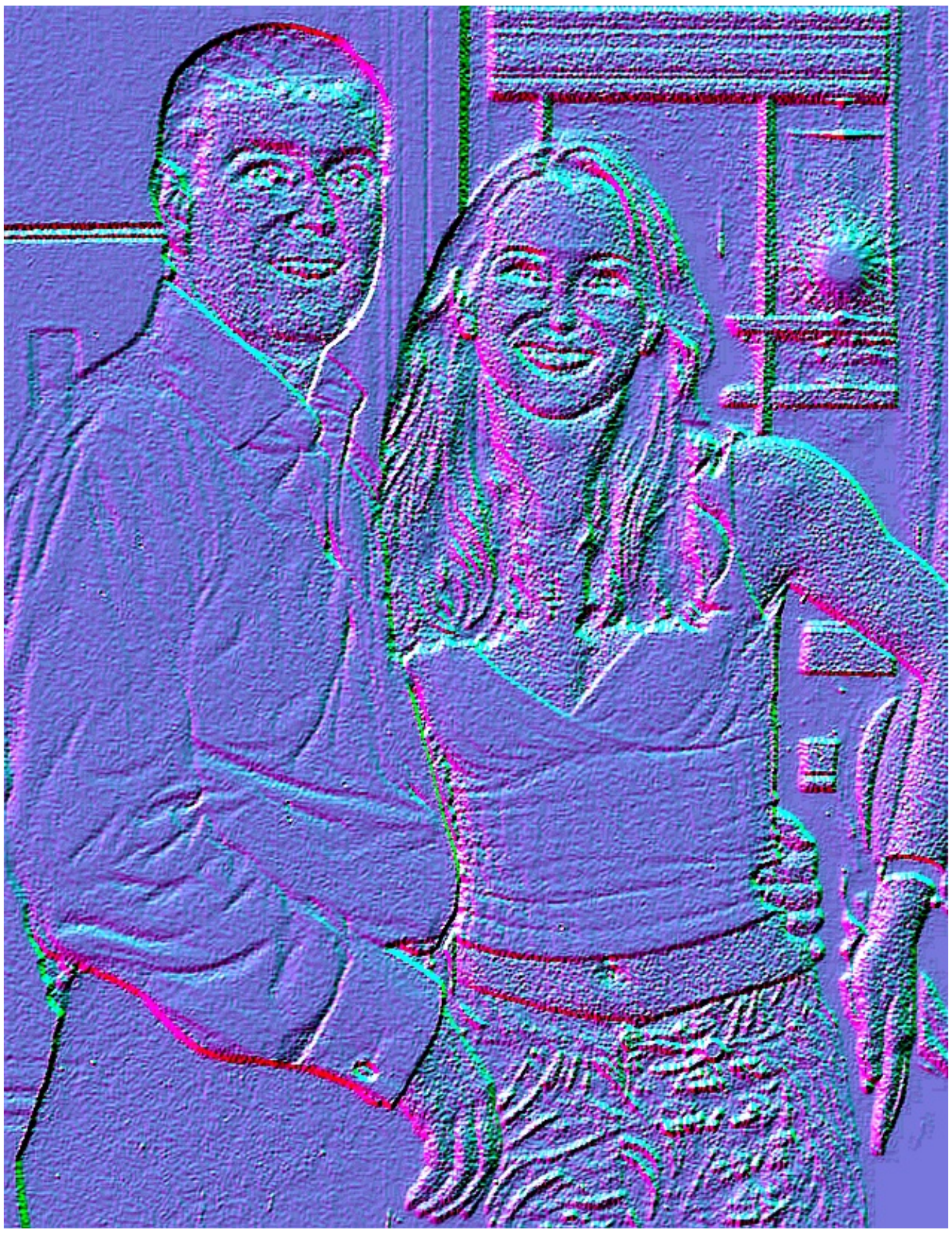

Fig. 9: TruFor forgery localization output of DM2023. Red indicates the highest probability of manipulation, grading through yellow and green to blue for unaltered regions. The confidence map (not shown) masks areas where predictions may be unreliable.

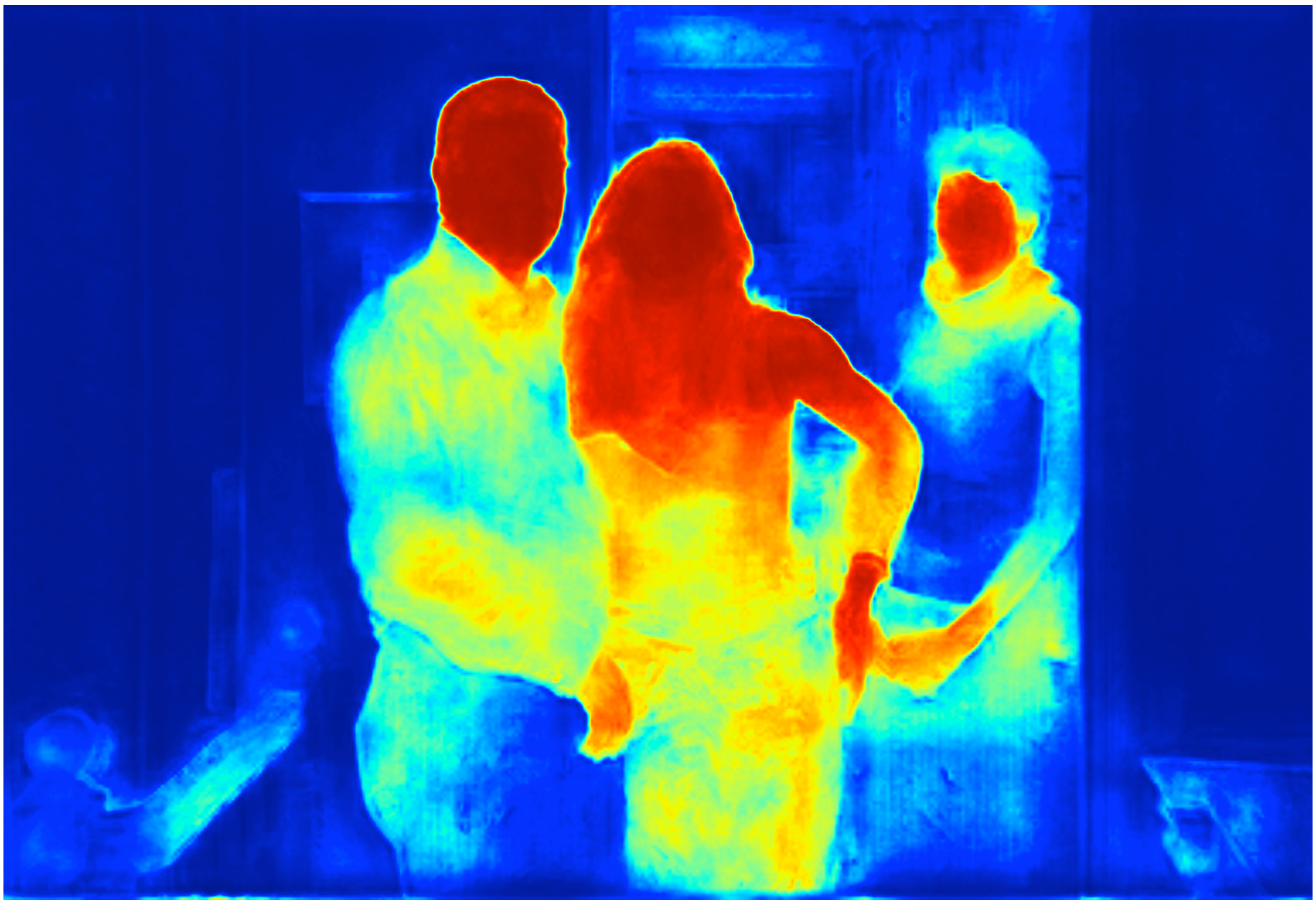

Fig. 10. Image crop from the "102 Dalmatians" reception 6th December 2000 (left) and DM2023 (right). Top row: native images; note the edge flare along the neckline. Middle row: size-matched crops with arrows marking the minimal geometric distortions required to align the two images. Bottom row: specular highlights from frontal flash illumination.

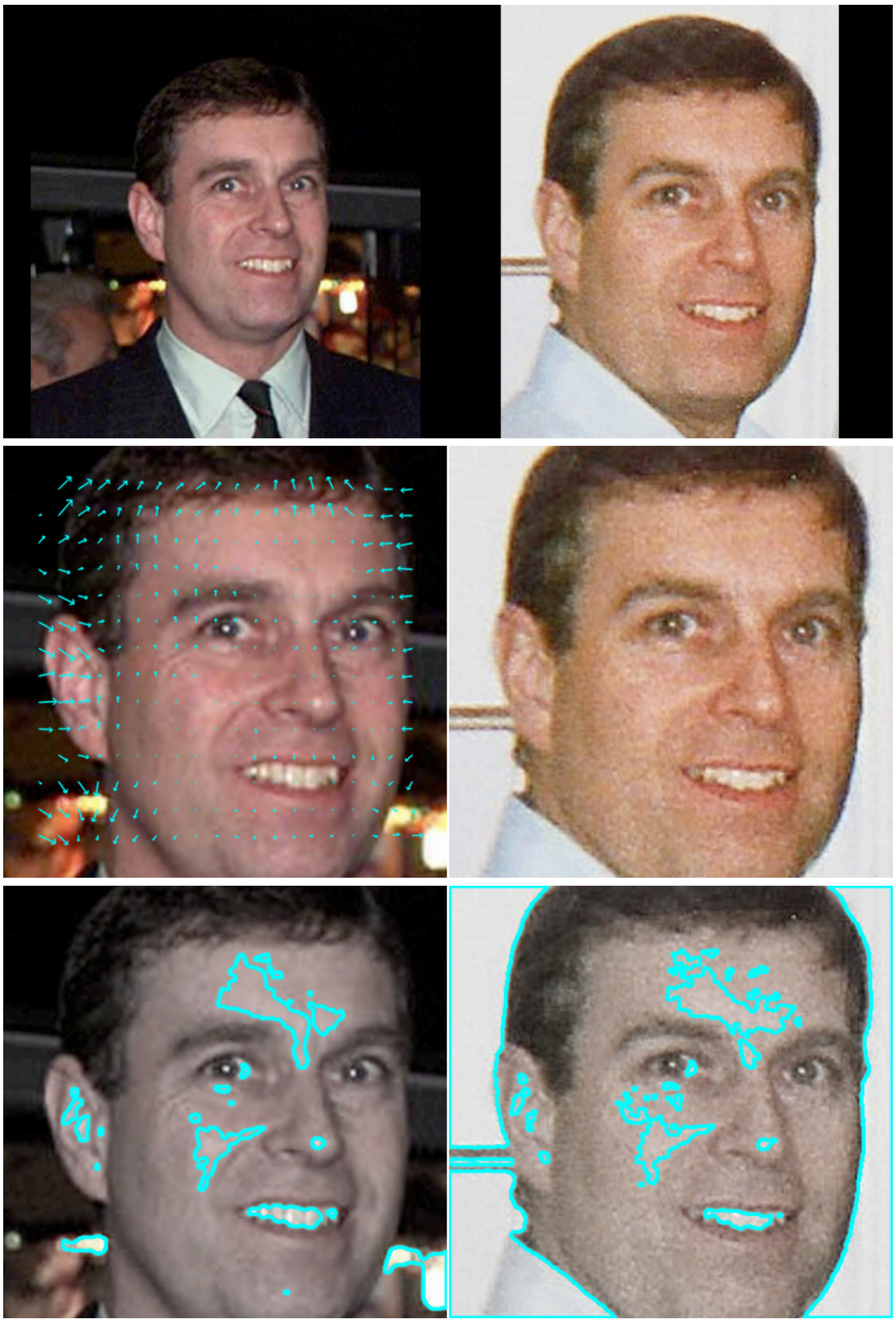

Fig. 11. Image crop from the "Anand Jon Fashion Show" in New York 18th September 2000 (left) and DM2023 (right). Top row: native images. Middle row: size-matched crops with arrows marking the minimal geometric distortions. Bottom row: specular highlights from frontal flash illumination.

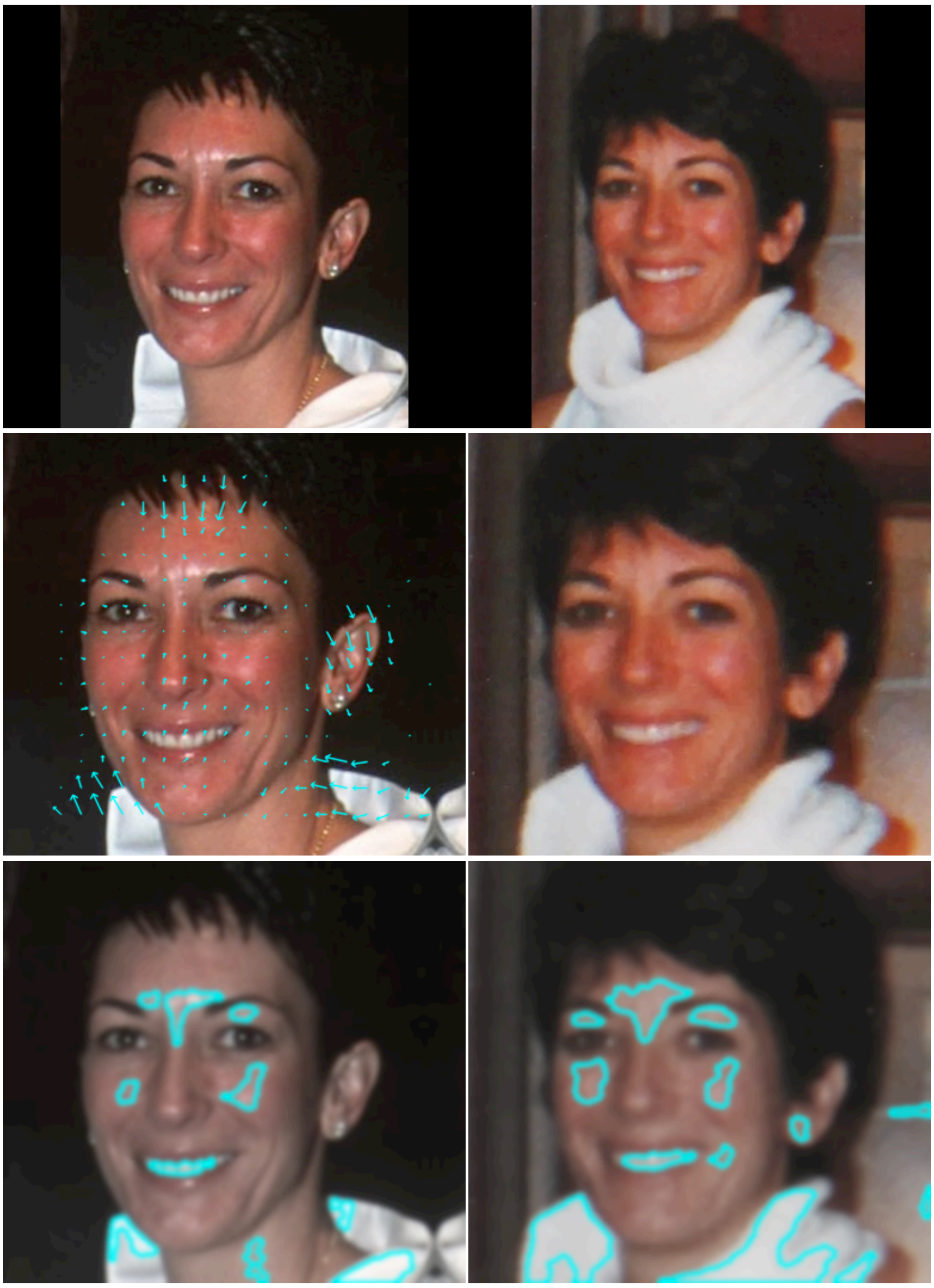

## Supplemental Files

Supporting files can be accessed at https://github.com/under-score/little_prince

## Acknowledgment

I wish to thank Drago N. for programming the Blender scene. The analysis would not have been possible without Guido Bartoli who developed Sherloq.

## Funding

None

## Ethics / Data availability

Only public available data have been used where all sources are given.

## Conflicts of Interest

The author declares no conflicts of interest.

He contacted Hany Farid May 5, 2025 by email for collaboration but he declined for being over-committed. On June 21, 2025 he contacted photographer Michael Thomas for the raw file of his image reproduction which was declined. On June 22, 2025 he asked the anonymous X account @RealMotherFaker for further details without any response. On August 1, 2025, he emailed Mail on Sunday reporter Mark Hookham who did not respond. On April 20, 2026 an email signed by "cbrobbin" suggested the source of the Maxwell image..

Author contribution

Study design, analysis, writing. and proof reading by the author.